\documentclass[journal,onecolumn,11pt,english]{IEEEtran}

\usepackage{array}
\usepackage{amsthm}
\usepackage{amsmath}
\usepackage{amssymb}
\usepackage{multirow}
\usepackage{flushend}
\usepackage{graphicx}
\usepackage[T1]{fontenc}
\usepackage[latin9]{inputenc}

\theoremstyle{plain}
\newtheorem{thm}{\protect\theoremname}
\theoremstyle{plain}
\newtheorem{lem}[thm]{\protect\lemmaname}
\theoremstyle{plain}
\newtheorem{cor}[thm]{\protect\corollaryname}

\usepackage{babel}
\providecommand{\lemmaname}{Lemma}
\providecommand{\theoremname}{Theorem}
\providecommand{\corollaryname}{Corollary}

\newcounter{example}
\newenvironment{example}[1][]{\refstepcounter{example}\par\medskip
\noindent \textit{Example~\theexample. #1} \rmfamily}{\medskip}

\makeatother

\newcommand{\cB}{\mathcal{B}}
\newcommand{\cI}{\mathcal{I}}
\newcommand{\cC}{\mathcal{C}}

\newcommand{\cK}{\mathcal{K}}

\begin{document}

\title{On the Duality and File Size Hierarchy of Fractional Repetition Codes}

\author{Bing~Zhu, Kenneth~W.~Shum, and~Hui~Li

\thanks{B. Zhu is with the School of Electronic and Computer Engineering, Peking University, P. R. China (e-mail: zhubing@sz.pku.edu.cn).}
\thanks{K. W. Shum is with the Institute of Network Coding, The Chinese University of Hong Kong (e-mail: wkshum@inc.cuhk.edu.hk).}
\thanks{H. Li is with the Shenzhen Key Lab of Information Theory and Future Network Architecture and the Future Network PKU Lab of National Major Research Infrastructure, Peking University Shenzhen Graduate School, Shenzhen 518055, P. R. China (e-mail: lih64@pkusz.edu.cn).}}

\maketitle

\begin{abstract}
Distributed storage systems that deploy erasure codes can provide better features such as lower storage overhead and higher data reliability. In this paper, we focus on fractional repetition (FR) codes, which are a class of storage codes characterized by the features of uncoded exact repair and minimum repair bandwidth. We study the \textit{duality} of FR codes, and investigate the relationship between the supported file size of an FR code and its dual code. Based on the established relationship, we derive an improved dual bound on the supported file size of FR codes. We further show that FR codes constructed from $t$-designs are optimal when the size of the stored file is sufficiently large. Moreover, we present the tensor product technique for combining FR codes, and elaborate on the file size hierarchy of resulting codes.
\end{abstract}


\section{Introduction}

Modern distributed storage systems are often built on thousands of inexpensive servers and disk drives. In such an architecture, data objects are fragmented and spread across a massive collection of physically independent storage devices (e.g., Google file system~\cite{key-1} and Hadoop distributed file system~\cite{key-2}). However, due to the commodity nature of practical data storage servers, component failures are prevalent in real-world storage environments~\cite{key-3,key-4}. To provide high reliability and availability, data redundancy should be employed in distributed storage systems.

Replication-based strategy is the simplest method to provide fault tolerance against failures~\cite{key-1,key-2}, where several copies of each data object are created and arranged on different storage nodes. Although data replication is easy to implement and manage, it suffers from the drawback of low storage efficiency. For the same level of redundancy, erasure coding technique can improve data reliability as compared to the replication scheme~\cite{key-5}. Maximum-distance-separable (MDS) codes are a class of erasure codes capable of providing the optimal trade-off between redundancy and reliability. In an erasure code based system, any data collector is able to reconstruct the original data file by contacting a certain number of nodes in the system. Upon failure of a node, the lost data should be recovered and stored in a replacement node by connecting to some surviving nodes (called \textit{helpers}) in this system. Even though traditional erasure codes can save the storage space, they generally require the retrieval of large amounts of data downloaded from helpers when repairing a single failed node. For example, an $[n,k]$ MDS code encodes a data object of $k$ fragments into $n$ storage nodes such that any subset of $k$ nodes are eligible for data retrieval. However, the system needs to recover the entire file in order to repair a node failure, which thus results in a large consumption of network resources (e.g., disk read and network transfer).

Regenerating codes are a class of erasure codes proposed in~\cite{key-6} with the capability to minimize the bandwidth consumption during the repair process. An $(n,k,d,\alpha,\beta)$ regenerating code encodes a data file into $n\alpha$ coded packets, which are spread across a storage system consisting of $n$ nodes, each having a capacity of $\alpha$. The stored file can be recovered by downloading data from any $k$ storage nodes in the system. When a node fails, the lost coded packets can be regenerated by connecting to any set of $d\geq k$ surviving nodes and~downloading $\beta$ packets from each node with a total repair bandwidth of $d\beta$. In particular, minimum-bandwidth regenerating (MBR) codes can recreate a failed node with the minimum repair bandwidth, i.e., $d\beta=\alpha$. We refer the readers to~\cite{key-7}\textendash{}\cite{key-9} for explicit constructions of regenerating codes.

Although MBR codes enjoy the minimum repair bandwidth, they impose an additional encoding complexity into the helper nodes contacted in the repair process. Specifically, each helper node needs to read all the packets it stored and transfer a linear combination of the retrieved data, which entails a large number of computations and disk read operations. Motivated by this, a simplified repair scheme called \textit{repair-by-transfer}, is presented in~\cite{key-7}, wherein the lost packets are recovered by duplicating the copies from some surviving nodes. Subsequently, El Rouayheb and Ramchandran~\cite{key-10} generalized the code constructions of~\cite{key-7} and introduced a new class of codes, termed fractional repetition (FR) codes, in which a two-layer encoding structure is employed to ensure data reconstruction and low-complexity node repair. The data objects are encoded in the first layer by an MDS code, and then the coded packets are replicated and stored in the~system according to the FR code in the second layer. In the presence of node failures, each helper node transfers a portion of stored data to the replacement node without performing additional encoding operations. By storing the transferred data, the replacement node maintains the same content as in the failed node. In such a sophisticated manner, FR codes enable \textit{uncoded} exact repairs at the MBR point. However, in contrast to traditional MBR codes, the node repair process of FR codes is table-based, which indicates that the failed node can be regenerated by contacting some specific subsets of surviving nodes~\cite{key-10}.

The capacity of a distributed storage system is the maximum amount of data that can be delivered to a data collector when contacting any $k$ out of $n$ storage nodes in the system~\cite{key-6}. The parameter $k$ is called the \textit{reconstruction degree}. In~\cite{key-6}, Dimakis~\textit{et al.}~theoretically showed that the storage capacity of an $(n,k,d,\alpha,\beta)$ MBR code based system is
\begin{equation}
\Big[kd-\binom{k}{2}\Big]\beta.
\label{Cap_MBR}
\end{equation}

Due to the different requirements in the node repair process, the MBR capacity given in \eqref{Cap_MBR} is not applicable to FR codes. For example, the FR codes constructed in~\cite{key-10} have a capacity greater than~or equal to that of MBR codes for $k\leq \alpha$. Indeed, the data reconstruction mechanism of FR codes is built on the outer MDS code. The supported file size%
\footnote{We notice that the supported file size of a given FR code is equivalent to~the storage capacity of the FR code based system.}
of an FR code essentially equals to the number of guaranteed distinct packets when downloading data from any collection of $k$ nodes. Intuitively, we can obtain the file size of a certain FR code by exhaustively considering all the $\binom{n}{k}$ possible combinations of $k$ nodes in the system. However, the computational complexity increases as $n$ and $k$ increase. On the other hand, having a knowledge of the supported file size is critical to the design of FR codes, which can be set as the input size of the outer MDS code.

\subsection{Related Work}

The concept of an FR code is introduced in the pioneer work~\cite{key-10}, wherein the authors also proposed explicit code constructions from regular graphs and Steiner systems. Several recent studies extend the construction of FR codes to a larger set of parameters, which are mainly based on the graph theory (e.g., bipartite cage graph~\cite{key-11} and extremal graph~\cite{key-12,key-13}) and combinatorial designs (e.g., transversal designs~\cite{key-12}, resolvable designs~\cite{key-14}, group divisible designs~\cite{key-15}, Hadamard designs~\cite{key-16}, perfect difference families~\cite{key-17}, relative difference sets~\cite{key-18} and partially ordered sets \cite{key-19}). Further, Pawar~\textit{et al.}~\cite{key-20} proposed a randomized scheme for constructing FR codes, which is based on the balls-and-bins model. In~\cite{key-21}, Anil~\textit{et al.} presented an incidence matrix based algorithm for designing FR codes, where they also enumerated FR codes up to a given number of nodes. Constructions of FR codes for dynamic data storage systems are considered in~\cite{key-22,key-23}, where the code parameters can evolve over time. The authors in \cite{key-24}\textendash{}\cite{key-26} investigated the constructions of FR codes with small repair degrees ($d < k$). Moreover, generalization of FR codes to heterogeneous storage networks is discussed in \cite{key-27}\textendash{}\cite{key-31}, where the system nodes have different storage capacities.

In addition to code constructions, some upper bounds on the maximum supported file size of FR codes with given parameters are also investigated in~\cite{key-10,key-12,key-16}. El Rouayheb and Ramchandran provided in~\cite{key-10} two upper bounds on the file size of FR codes. Subsequently, Silberstein and Etzion presented in~\cite{key-12} explicit code constructions that attain these bounds. Furthermore, Olmez and Ramamoorthy determined the supported file size for most of their code constructions~\cite{key-16}.

\subsection{Our Contributions}

In this paper, we investigate the duality of FR codes, and establish a close relationship between the supported file size of an FR code and its dual code. Specifically, our main contributions are three-fold.

\begin{enumerate}
\item By jointly considering the relationship and the upper bound in~\cite{key-10}, we provide an improved upper bound on the supported file size of FR codes, which is referred to as the \textit{dual bound}.
\item From the dual perspective, we show that FR codes based on $t$-designs are optimal when the size of the stored file is sufficiently large.
\item We present the tensor product method for combining two FR codes. The file size hierarchy of the resulting code can be obtained from those of the component codes.
\end{enumerate}

The rest of this paper is organized as follows. Section II introduces the necessary background and notations. Section III provides a dual bound on the supported file~size of FR codes. Section IV shows that FR codes derived from $t$-designs are optimal for certain parameter ranges. Section V discusses the tensor product of FR codes. Finally, Section VI concludes the paper.

\section{Preliminaries}

\subsection{Incidence Structure and $t$-Designs}

An \textit{incidence structure} is a triple $(P,\cB,\cI)$, where $P$ and $\cB$ are nonempty finite sets, and $\cI$ is a subset of $P\times\cB$. The elements in $P$ are called \textit{points}, and the elements in $\cB$ are called \textit{blocks}. An element in $\cI$ is called a {\em flag}, and we say that a point $p\in P$ is \textit{incident} with a block $B\in \cB$ if $(p,B)$ is a flag in $\cI$. We can also specify an incidence structure by an \textit{incidence matrix}, which is a $|\cB|\times|P|$ zero-one matrix with rows indexed by the blocks and columns indexed by the points, such that the entry corresponding to a point $p$ and a block $B$ is equal to $1$ if and only if $p$ is incident with $B$. If an incidence matrix has constant row sums and constant column sums, then the corresponding incidence structure is called a {\em tactical configuration}~\cite{key-32}.

In this general setting, it is permissible that two distinct blocks are incident with the same set of points, and if it occurs, we say that there are \textit{repeated blocks}. An incidence structure with no repeated blocks is called {\em simple}. In a simple incidence structure, we can identify a block with a subset of $P$, and denote the incidence structure by $(P,\cB)$.

A \textit{$t$-design} is a simple incidence structure in which every block has the same size and any $t$ distinct points are contained in exactly $\lambda$ blocks, for some constants $t$ and $\lambda$. More precisely, for positive integers $t$, $m$, $\lambda$, and $v$ satisfying $t\leq m<v$, a $t$-$(v,m,\lambda)$ {\em design} is a simple incidence structure $(P,\cB)$ such that (i) $|P| = v$, (ii) $|B| =m$ for all $B\in \cB$, and (iii) any subset of $t$ points of $P$ occurs in exactly $\lambda$ blocks in $\cB$. When $t=1$, a $1$-design is nothing but a simple tactical configuration.

For example, consider a point set $P=\{1,2,\ldots,7\}$ and a block set $\cB=\{\{1,2,3,6\},\{1,2,5,7\},\{1,3,4,5\},\{1,$ $4,6,7\},\{2,3,4,7\},\{2,4,5,6\},\{3,5,6,7\}\}$. We note that every pair of points appears in exactly two blocks. Thus, $(P,\cB)$ forms a $2$-$(7,4,2)$ design.

\begin{lem}(\cite[Theorem 9.7]{key-33})
\label{Basic}
Suppose that $(P,\cB)$ is a $t$-$(v,m,\lambda)$ design. Let $X$ and $Y$ be disjoint subsets of $P$ such that $|X|=i$, $|Y|=j$, and $i+j\leq t$. Then, there are exactly
\begin{equation}
\label{Count}
\lambda^j_i := \lambda\frac{\tbinom{v-i-j}{m-i}}{{\tbinom{v-t}{m-t}}}
\end{equation}
blocks in $\cB$ that contain all the points in $X$ and none of the points in $Y$.
\end{lem}

For the special case that $i=j=0$, we obtain the number of blocks in a $t$-$(v,m,\lambda)$ design, which is given by
\begin{equation}
b := \lambda^0_0=\lambda\frac{\tbinom{v}{t}}{{\tbinom{m}{t}}}.
\end{equation}
Moreover, if $|X|=1$ and $|Y|=0$, we have $\lambda^0_1=\lambda\tbinom{v-1}{m-1}/{\tbinom{v-t}{m-t}}$, implying that each point is contained in $\lambda^0_1$ blocks.

\subsection{DRESS Code and Fractional Repetition Code}

A {\em Distributed Replication-based Exact Simple Storage} (DRESS) code is a coding architecture that consists of an outer code and an inner code described as follows~\cite{key-10}. The outer code is an MDS code with dimension $M$ and length $\theta$ over a sufficiently large finite field. To distribute a data object of size $M$, which is referred to as a \textit{data file}, we first encode it by the outer $[\theta,M]$ MDS code, such that any $M$ out of the obtained $\theta$ coded packets are sufficient to reconstruct the data file. In the following, we will use symbols and packets interchangeably. The inner code is an incidence structure $\cC = (P,\cB,\cI)$ such that the symbols produced by the outer MDS code are indexed by the points in $P$ (i.e., $|P| = \theta$). Each storage node is associated with a unique block in $(P,\cB,\cI)$, and stores the coded symbols indexed by the points in the corresponding block.

For a given reconstruction degree $k$, the {\em supported file size} of the inner code $\cC = (P,\cB,\cI)$ is defined as
\begin{equation}
M_k(\cC) := \min_{\mathcal{K}\subset \cB, |\mathcal{K}|=k} |\{p\in P:\, \exists B \in \mathcal{K}, (p, B) \in \cI \}|,
\label{eq:M}
\end{equation}
where the minimum is taken over all $k$-subsets $\cK$ of the block set $\cB$. By definition, the value of $M_k(\cC)$ refers to the number of guaranteed distinct packets one can download from any $k$ storage nodes. For a fixed value of $k$, we can choose an outer MDS code with length $|P|$ and dimension $M_k(\cC)$, such that any subset of $k$ nodes are sufficient in decoding the data object.

The design rationale of the inner code is to facilitate node repair. Upon failure of a storage node, each helper node simply passes the packets it has in common with the failed node for repair. In other words, DRESS codes enjoy the repair efficiency of the replication scheme, and are suitable for high-churn environments with frequent node joins/leaves (e.g., peer-to-peer distributed storage systems). Friedman~\textit{et al.}~\cite{key-34} evaluated the efficiency of DRESS codes in practical peer-to-peer environments, and showed that the concatenated scheme can achieve better features than each of the methods separately. Moreover, Itani~\textit{et al.}~\cite{key-35,key-36} investigated the optimal repair cost of DRESS code based data storage systems, where they proposed efficient genetic algorithms for the single node failure and multiple node failure scenarios respectively.

In this paper, we concentrate on DRESS codes which employ a tactical configuration as the inner code. We define a {\em fractional repetition} (FR) code as a tactical configuration $(P,\cB,\cI)$ with $\theta$ points and $n$ blocks, in which every point is incident with $\rho$ blocks, and every block is incident with $\alpha$ points, for some constants $\rho$ and $\alpha$. Hence, every coded packet is replicated $\rho$ times in the storage system, and each storage node contains $\alpha$ packets. We refer to such an FR code as an $(n, \alpha, \theta, \rho)$-FR code, and call the parameter $\rho$ the {\em repetition degree}.

Since the incidence matrix of an FR code has constant row sum $\alpha$ and constant column sum~$\rho$, we have the following basic relation
\begin{equation}
\label{eq:basic}
n\alpha = \theta\rho
\end{equation}
among the code parameters.

\begin{figure}
\centering{}\includegraphics[scale=0.125]{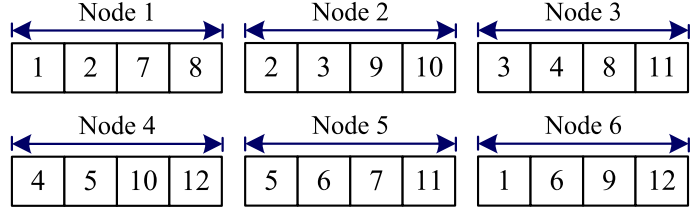}
\caption{An FR code with repetition degree $\rho=2$ for a distributed storage system with $n=6$ nodes. The numbers in the blocks give the indices~of packets stored in the node, i.e., each storage node contains $\alpha=4$ packets.}
\label{Example}
\end{figure}

We illustrate how to distribute data packets across a storage system using the $(6,4,12,2)$-FR code shown in Fig.~\ref{Example}. By using a $[12,9]$ MDS code as the outer code, we encode a data file consisting of $9$ source symbols to $12$ coded symbols. These coded symbols are then distributed to $6$ storage nodes according to the incidence structure in Fig.~\ref{Example}. Furthermore, we observe that a data collector contacting any $3$ nodes can obtain at least $9$ distinct coded packets, which are sufficient to decode the original data.

Suppose that $\cC = (P,\cB,\cI)$ is an FR code. The \textit{dual} of $\cC$ is defined as the FR code $(\cB,P,\cI^t)$, where  $\cI^t$ is the subset of $\cB\times P$ defined by
$$
\cI^t := \{(B,p):\, (p,B)\in \cI\}.
$$
We denote the dual of $\cC$ by $\cC^t$. Notice that the incidence matrix of $\cC$ and $\cC^t$ are the transpose of each other. In \cite{key-10}, the authors refer to the dual FR code as the \textit{transpose code}. We state two immediate properties of dual codes in the following lemma.

\begin{lem}
Let $\cC$ be an $(n,\alpha, \theta, \rho)$-FR code.

(i) The dual code of $\cC$ is a $(\theta, \rho, n, \alpha)$-FR code.

(ii) The double dual of $\cC$ is $\cC$ itself.
\label{lemma:easy}
\end{lem}

\begin{example}
The ``repetition code'' on $n$ storage nodes is an $(n,1,1,n)$-FR code. The incidence matrix is the all-one $n\times 1$ matrix. By definition, the dual of this repetition code is a $(1,n,n,1)$-FR code, which consists of a single storage node containing all the coded symbols.
\end{example}

\section{The Hierarchy of Supported File Size and The Dual Bound}

\subsection{File Size Hierarchy of FR Codes}

Suppose that $\cC = (P,\cB,\cI)$ is an $(n,\alpha, \theta, \rho)$-FR code. The supported file size $M_k(\cC)$ of $\cC$ is a non-decreasing function of $k$, i.e.,
\begin{equation}
\alpha = M_1(\cC) \leq M_2(\cC) \leq \cdots \leq M_n(\cC)=\theta.
\label{eq:chain1}
\end{equation}
We call the above the \textit{hierarchy of supported file size} of $\cC$.%
\footnote{In~\cite{key-12}, the authors introduced the notion of file size hierarchy of FR codes for $1\leq k \leq \alpha$. We extend this study in this paper by taking all the possible reconstruction degrees into consideration.}
We also define $M_0(\cC):=0$ by convention. Similarly, the file size hierarchy of the dual code $\cC^t$ is
\begin{equation}
\rho = M_1(\cC^t) \leq M_2(\cC^t) \leq \cdots \leq M_\theta(\cC^t)=n.
\label{eq:chain2}
\end{equation}

Note that there is a close relationship between $M_k(\cC)$ and $M_\ell(\cC^t)$. This property can be seen from the fact that if we~can find an $x\times y$ all-zero submatrix in the incidence matrix of $\cC$, then we have
\begin{equation}
M_x(\cC) \leq  \theta-y,
\end{equation}
and
\begin{equation}
M_y(\cC^t) \leq n-x.
\end{equation}

This motivates us to define
$$
N_k(\cC) := |P| - M_k(\cC)
$$
\begin{equation}
= \max_{\mathcal{K}\subset \cB, |\mathcal{K}|=k} |\{p\in P: \not\exists B\in \mathcal{K}, (p,B)\in \mathcal{I}\}|
\label{NkC}
\end{equation}
with the maximum taken over all subsets $\mathcal{K}\subset \cB$ of size $k$. By definition, $N_k(\cC)$ is the largest integer $\ell$ such that we can find a $k \times \ell$ all-zero submatrix in the incidence matrix of $\cC$. From \eqref{eq:chain1} and \eqref{eq:chain2}, we have
\begin{gather*}
\theta = N_0(\cC) > N_1(\cC) \geq N_2(\cC) \geq \cdots \geq N_n(\cC) = 0, \text{ and } \\
n = N_0(\cC^t) > N_1(\cC^t) \geq N_2(\cC^t) \geq \cdots \geq N_\theta(\cC^t) = 0.
\end{gather*}

The following result follows directly from the relation of $N_k(\cC)$ and $N_\ell(\cC^t)$, where $0 \leq k \leq n$ and $0 \leq \ell \leq \theta$.

\begin{lem}
Let $\cC$ be an FR code and let $k_0$ be a given reconstruction degree. Denote $N_{k_0}(\cC)$ as $\ell_0$ and $N_{\ell_0}(\cC^t)$ as $k_1$. Then, we have (i) $k_1 \geq k_0$, and (ii) $N_{k_1}(C) = \ell_0$.
\end{lem}

\begin{figure}
\centering{}\includegraphics[scale=0.11]{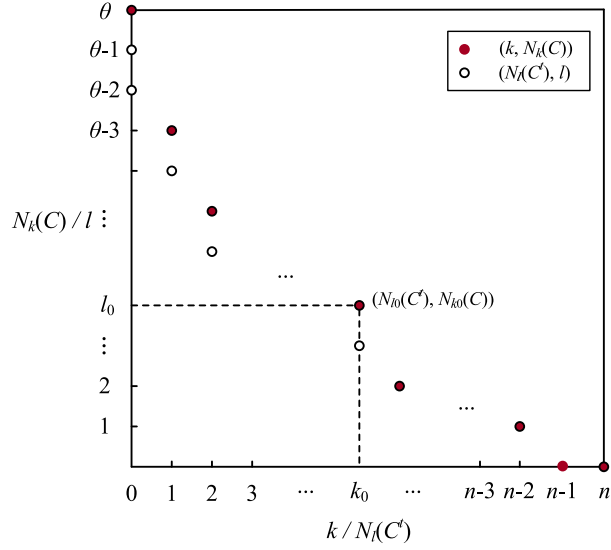}
\caption{The graphic illustration of $(k,N_k(\cC))$ and $(N_\ell(\cC^t),\ell)$.}
\label{Pareto}
\end{figure}

We now plot the points $(k,N_k(\cC))$ for $k=0,1,\ldots, n$, and $(N_\ell(\cC^t),\ell)$ for $\ell=0,1,\ldots, \theta$ in the same figure.~The results can be found in Fig.~\ref{Pareto}. A Pareto optimal point, say $(k_0,\ell_0)$,~is a vertex of the graph that satisfies
$$
 \ell_0 = N_{k_0}(\cC) \text{ and } k_0 = N_{\ell_0}(\cC^t),
$$
and
\begin{gather*}
N_k(\cC) < N_{k_0}(\cC) \text{ for all } k > k_0, \\
N_\ell(\cC^t) < N_{\ell_0}(\cC^t) \text{ for all } \ell > \ell_0.
\end{gather*}
Therefore, we obtain
\begin{equation}
N_k(\cC) = \begin{cases}
\theta, & \text{for } k=0, \\
\theta-1, & \text{for } 0 = N_\theta(\cC^t) < k \leq N_{\theta-1}(\cC^t), \\
\theta-2, & \text{for } N_{\theta-1}(\cC^t) < k \leq N_{\theta-2}(\cC^t), \\
\vdots & \vdots \\
1, & \text{for } N_2(\cC^t) < k \leq N_{1}(\cC^t), \\
0, & \text{for } N_1(\cC^t)< k \leq N_0(\cC^t) = n.
\end{cases}
\end{equation}

Based on the above analysis, we obtain the following theorem.
\begin{thm}
Let $\cC$ be an $(n,\alpha,\theta,\rho)$-FR code. With $N_\ell(\cC^t)$~as defined in \eqref{NkC}, we have
\begin{equation}
M_k(\cC) = \begin{cases}
\theta, & \text{for } N_1(\cC^t) <k \leq n = N_0(\cC^t), \\
\theta-1, & \text{for } N_2(\cC^t) <k \leq  N_1(\cC^t), \\
\theta-2, & \text{for } N_3(\cC^t) <k \leq  N_2(\cC^t), \\
\vdots & \vdots \\
2, & \text{for } N_{\theta-1}(\cC^t) <k \leq  N_{\theta-2}(\cC^t),\\
1, & \text{for } N_\theta(\cC^t)=0 <k \leq  N_{\theta-1}(\cC^t).
\end{cases}
\label{eq:Mk}
\end{equation}
\label{thm:duality}
\end{thm}

\textit{Remark 1.} We notice that the identities in \eqref{eq:Mk} can be expressed in a more compact way by
\begin{equation}
M_k(\cC) = \sum_{i=1}^\theta \mathbb{I}(k > N_i(\cC^t)),
\label{eq:indicator}
\end{equation}
where $\mathbb{I}(C)$ is the indicator function equal to $1$ if the condition $C$ is true and $0$ otherwise. In this case, the right-hand side term of \eqref{eq:indicator} counts the number of $i\in\{1,2,\ldots, \theta\}$ such that $N_i(\cC^t)$~is strictly less than $k$. Thus,
$$
\sum_{i=1}^\theta \mathbb{I}( k > N_i(\cC^t)) = \theta-\ell \text{ for } N_{\ell+1}(\cC^t) < k \leq N_\ell(\cC^t),
$$
where $k=1,2,\ldots, n$.

\begin{example}
Let $\cC$ be the incidence structure obtained from the line graph of the complete graph on five vertices. This gives the $(5,4,10,2)$-FR code with incidence matrix
$$
{\begin{bmatrix}
1 & 1 & 1 & 1 & 0 & 0 & 0 & 0 & 0 & 0\\
1 & 0 & 0 & 0 & 1 & 1 & 1 & 0 & 0 & 0\\
0 & 1 & 0 & 0 & 1 & 0 & 0 & 1 & 1 & 0\\
0 & 0 & 1 & 0 & 0 & 1 & 0 & 1 & 0 & 1\\
0 & 0 & 0 & 1 & 0 & 0 & 1 & 0 & 1 & 1
\end{bmatrix}
}
$$
as discussed in \cite{key-10}. This is a $5\times 10$ matrix with constant row sum $\alpha=4$ and constant column sum $\rho=2$. The $5$ blocks in this FR code are
$$
\{1,2,3,4\}, \{1,5,6,7\}, \{2,5,8,9\}, \{3,6,8,10\},  \{4,7,9,10\}.
$$

\begin{figure}
\centering{}\includegraphics[scale=0.105]{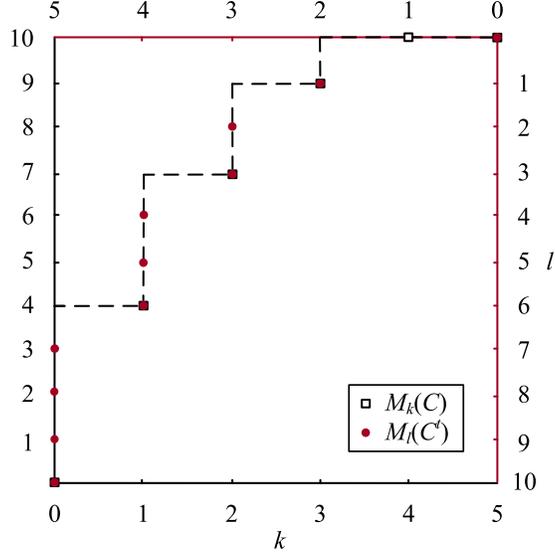}
\caption{The relationship between $M_k(\cC)$ and $M_\ell(\cC^t)$.}
\label{Example2}
\end{figure}

For $k=1,\ldots, 5$, we can compute that the supported file size $M_k(\cC)$ of the complete graph based FR code $\cC$ is
$$
M_k(\cC) = \begin{cases}
10, & \text{for }k= 4, 5, \\
9, & \text{for } k = 3, \\
7, & \text{for } k = 2, \\
4, & \text{for } k = 1,
\end{cases}
$$
and the values of $N_\ell(\cC^t)$ for $\ell=1,\ldots,10$ are
\begin{align*}
0& =N_7(\cC^t) = N_8(\cC^t)=N_9(\cC^t)=N_{10}(\cC^t), \\
1&= N_4(\cC^t)=N_5(\cC^t)=N_6(\cC^t), \\
2& =N_2(\cC^t) = N_3(\cC^t), \\
3&=N_1(\cC^t).
\end{align*}

Moreover, the supported file size hierarchy of $\cC^t$ is $5-N_\ell(\cC^t)$, i.e,
\begin{align*}
5 &= M_{10}(\cC^t) = M_{9}(\cC^t) =M_{8}(\cC^t) = M_{7}(\cC^t), \\
4 &= M_{6}(\cC^t) = M_{5}(\cC^t) =M_{4}(\cC^t), \\
3 &= M_{3}(\cC^t) = M_{2}(\cC^t),\\
2 &= M_{1}(\cC^t).
\end{align*}
Fig.~\ref{Example2} illustrates the relationship between $M_k(\cC)$ and $M_\ell(\cC^t)$. We can obtain the two supported file size functions if we view the stair-case graph from two different perspectives, which are distinguished with different colors.
\end{example}

\subsection{An Improved Dual Bound}

In~\cite{key-10}, the authors showed that the supported file size $M_k(\cC)$~of an $(n,\alpha,\theta,\rho)$-FR code $\cC$ is upper bounded by
\begin{equation}
M_k(\cC) \leq g(k),
\label{eq:bound2}
\end{equation}
where $g(k)$ is defined recursively by
$$
g(1) := \alpha, \ g(k+1) := g(k) + \alpha - \Big\lceil \frac{\rho g(k) - k \alpha}{n-k}\Big\rceil.
$$

Note that Theorem~\ref{thm:duality} provides a link between an FR code and its dual. Using the mechanism in the previous subsection, we can obtain an improved upper bound if we take the upper bound in~\eqref{eq:bound2} into consideration.

\begin{thm}
Given an FR code $\cC$ with parameters $(n,\alpha,\theta,\rho)$, we define the function $g'(\ell)$ recursively by
$$
g'(1) := \rho,\ g'(\ell+1) := g'(\ell) + \rho - \Big\lceil \frac{\alpha g'(\ell) - \ell \rho}{\theta-\ell}\Big\rceil.
$$
for $\ell=1,2,\ldots, \theta-1$. Then, for all $k = 1,2,\ldots, n$, we have
\begin{equation}
M_k(\cC) \leq \sum_{\ell=1}^\theta \mathbb{I}( k > n-g'(\ell)).
\label{eq:bound3}
\end{equation}
\end{thm}

\begin{IEEEproof}
The function $g'(\ell)$ is the counterpart of the recursive bound \eqref{eq:bound2} on the dual code. Thus,
\begin{equation}
M_\ell(\cC^t) \leq g'(\ell).
\end{equation}
Since
\begin{equation}
N_\ell(\cC^t) \geq n - g'(\ell),
\end{equation}
for all $\ell$, in view of the remark after Theorem~\ref{thm:duality}, we have
\begin{equation}
M_k(\cC) = \sum_{\ell=1}^\theta \mathbb{I}(k > N_\ell(\cC^t)) \leq \sum_{\ell=1}^\theta \mathbb{I}(k > n - g'(\ell)),
\end{equation}
which completes the proof.
\end{IEEEproof}

We refer to the inequality in \eqref{eq:bound3} as the \textit{dual bound} on the supported file size.

\begin{example} Consider an FR code $\cC$ with parameters $(n,\alpha,\theta,\rho) = (9,2,6,3)$. The bound in \eqref{eq:bound2} suggests that the supported file size with reconstruction degree $k=4$~is upper bounded by $M_4(\cC) \leq g(4) = 5$.

Moreover, the recursive bound applied to the dual code yields $M_\ell(\cC^t) \leq g'(\ell)$ with
\begin{gather*}
g'(1)=3,\  g'(2)=5,\ g'(3)=7, \\
g'(4)=8,\ g'(5)=g'(6)= 9.
\end{gather*}
Then, the dual bound in \eqref{eq:bound3} gives
$$
M_4(\cC) \leq \sum_{\ell=1}^{6} \mathbb{I}(4 > 9-g'(\ell)) = \sum_{\ell=1}^{6} \mathbb{I}(g'(\ell) > 5) = 4.
$$
This bound can be achieved by the $(9,2,6,3)$-FR code listed in the database \cite{key-21} with the following incidence matrix:
$$
{\left[\begin{array}{cccccc}
1& 1& 0& 0& 0& 0\\
1& 0& 1& 0& 0& 0\\
1& 0& 0& 1& 0& 0\\
0& 1& 0& 0& 1& 0\\
0& 1& 0& 0& 0& 1\\
0& 0& 1& 1& 0& 0\\
0& 0& 1& 0& 1& 0\\
0& 0& 0& 1& 0& 1\\
0& 0& 0& 0& 1& 1
\end{array} \right]
}.
$$

We observe that the four storage nodes associated to rows $1, 2, 3$ and $6$ contain precisely $4$ distinct packets. Thus, this FR code can support a file size of $M=4$ with $k=4$, implying that it is optimal by the dual bound.
\end{example}

\begin{table}
\protect\caption{Comparison Between the Recursive Bound in~\cite{key-10} and the Dual Bound}
\begin{centering}
\begin{tabular}{|c|c|c|c|}
\hline
Code Parameter & $k$ & Recursive Bound & Dual Bound\tabularnewline
\hline
\hline
$(10,2,5,4)$ & $3$ & $4$ & $3$\tabularnewline
\hline
$(10,4,10,4)$ & $4$ & $9$ & $8$\tabularnewline
\hline
$(10,4,8,5)$ & $3$ & $7$ & $6$\tabularnewline
\hline
$(11,3,11,3)$ & $6$ & $10$ & $9$\tabularnewline
\hline
$(11,4,11,4)$ & $5$ & $10$ & $9$\tabularnewline
\hline
\multirow{2}{*}{$(12,2,8,3)$} & $5$ & $6$ & $5$\tabularnewline
\cline{2-4}
 & $7$ & $7$ & $6$\tabularnewline
\hline
\multirow{2}{*}{$(12,2,6,4)$} & $3$ & $4$ & $3$\tabularnewline
\cline{2-4}
 & $5$ & $5$ & $4$\tabularnewline
\hline
$(12,3,12,3)$ & $7$ & $11$ & $10$\tabularnewline
\hline
$(12,4,12,4)$ & $6$ & $11$ & $10$\tabularnewline
\hline
$(12,5,15,4)$ & $6$ & $14$ & $13$\tabularnewline
\hline
$(12,6,18,4)$ & $6$ & $17$ & $16$\tabularnewline
\hline
$(12,7,21,4)$ & $6$ & $20$ & $19$\tabularnewline
\hline
$(12,8,24,4)$ & $6$ & $23$ & $22$\tabularnewline
\hline
$(13,3,13,3)$ & $8$ & $12$ & $11$\tabularnewline
\hline
$(13,8,26,4)$ & $7$ & $25$ & $24$\tabularnewline
\hline
$(14,8,28,4)$ & $8$ & $27$ & $26$\tabularnewline
\hline
$(14,12,42,4)$ & $8$ & $41$ & $40$\tabularnewline
\hline
\end{tabular}
\par\end{centering}
\label{Bound:Com}
\end{table}

Table~\ref{Bound:Com} shows the comparison between the recursive bound in~\cite{key-10} and our dual bound for some parameter ranges.

\section{Optimal FR Codes based on $t$-Designs}

Another upper bound on the supported file size $M_k(\cC)$ of an $(n,\alpha,\theta,\rho)$-FR code $\cC$ is derived in~\cite{key-10} as
\begin{equation}
M_k(\cC) \leq \Big\lfloor \theta \Big(1 - \frac{\binom{n-\rho}{k}}{\binom{n}{k}}\Big) \Big\rfloor.
\label{eq:bound1}
\end{equation}

From the dual perspective, we show that the bound in \eqref{eq:bound1} is essentially the same as the following bound on the reconstruction degree~$k$, which is first obtained in~\cite{key-12}.

\begin{lem} (\cite[Lemma 32]{key-12})
\label{thm:Silberstein}
If we store a data file of size $M$ by using an $(n,\alpha,\theta,\rho)$-FR code $\cC$, then the reconstruction degree $k$ is lower bounded by
\begin{equation}
k \geq \Big\lceil \frac{n \binom{M-1}{\alpha}}{\binom{\theta}{\alpha}}\Big\rceil + 1.
\label{eq:boundk}
\end{equation}
\end{lem}

\begin{IEEEproof}
By applying the bound in \eqref{eq:bound1} to the dual code of $\cC$, we obtain
\begin{equation}
M_\ell(\cC^t) \leq  n \Big(1 - \frac{\binom{\theta-\alpha}{\ell}}{\binom{\theta}{\ell}}\Big),
\end{equation}
for $\ell=1,2,\ldots, \theta$. (We can remove the floor operator without loss of generality.) Hence,
\begin{equation}
N_\ell(C^t) \geq n - n \Big(1 - \frac{\binom{\theta-\alpha}{\ell}}{\binom{\theta}{\ell}}\Big) = n \frac{\binom{\theta-\alpha}{\ell}}{\binom{\theta}{\ell}}.
\end{equation}

Given an integer $M$ between $1$ and $\theta$, we let $\ell$ be the integer that satisfies
$$
M = \theta -\ell+1.
$$
By Theorem~\ref{thm:duality}, we obtain
\begin{equation}
k \geq N_\ell(C^t) +1 \geq n \frac{\binom{\theta-\alpha}{\theta-M+1}}{\binom{\theta}{\theta-M+1}} +1 = n\frac{\binom{M-1}{\alpha}}{\binom{\theta}{\alpha}} + 1.
\end{equation}
The proof of this theorem is completed by taking the ceiling~of both sides.
\end{IEEEproof}

In what follows, we consider FR codes derived from $t$-designs. Recall that in a $t$-$(v,m,\lambda)$ design $(P,\cB)$, each point of $P$ is contained in the same number of $\lambda^0_1$ blocks. Therefore, we can obtain an FR code $\cC$ with repetition degree $\lambda^0_1$ by taking $\cC=(P,\cB)$.

We state the main result in the following theorem.

\begin{thm}
Let $(P,\cB)$ be a $t$-$(v,m,\lambda)$ design, and let $\cC$ be the FR code based on $(P,\cB)$. Then, the supported file size $M_k(\cC)$ is optimal for $k$ in the range $\lambda_{0}^t < k \leq \lambda_{0}^0 = b$, and is given by
\begin{equation}
M_k(\cC)=\left\{
\begin{array}{rcl}
v, & & \text{for }  {\lambda^{1}_0 < k \leq \lambda_{0}^0 = b},\\
v-1, & & \text{for }  {\lambda^{2}_0 < k \leq \lambda^{1}_0},\\
\vdots & & {\vdots}\\
v-t+2, & & \text{for }  {\lambda^{t-1}_0 < k \leq \lambda^{t-2}_0},\\
v-t+1, & & \text{for }  {\lambda^{t}_0 < k\leq \lambda^{t-1}_0}.
\end{array}
\right.
\end{equation}
\end{thm}

\begin{IEEEproof}
Let $L$ be an arbitrary $\ell$-sized subset of $P$, where $1\leq \ell \leq t$. Based on Lemma~\ref{Basic}, we obtain that the number of blocks in $\cB$ that are disjoint from $L$ is $\lambda^{\ell}_0$. Hence, for the constructed FR code $\cC$, we have
\begin{equation}
N_\ell(\cC^t)=\lambda^{\ell}_0=\lambda\frac{\tbinom{v-\ell}{m}}{\tbinom{v-t}{m-t}},
\label{compN}
\end{equation}
which in conjunction with Theorem~\ref{thm:duality} gives the file size of $\cC$.

Let $\ell'$ be an integer such that $0\leq \ell' \leq t-1$. By substituting $M=v-\ell'$ into \eqref{eq:boundk}, we obtain
$$
k \geq \Big\lceil \frac{b \binom{v-\ell'-1}{m}}{\binom{v}{m}}\Big\rceil + 1 = \Big\lceil \frac{\lambda\binom{v}{t}\binom{v-\ell'-1}{m}}{\binom{m}{t}\binom{v}{m}}\Big\rceil + 1
$$
\begin{equation}
=\frac{\lambda (v-\ell'-1)\cdots(v-\ell'-m)}{m(m-1)\cdots(m-t+1)(v-t)\cdots(v-m+1)}+1
\end{equation}
\begin{equation}
=\frac{\lambda \binom{v-\ell'-1}{m}}{\binom{v-t}{m-t}}+1 = \lambda^{\ell'+1}_0+1.
\end{equation}
Therefore, $\cC$ attains the lower bound in Lemma~\ref{thm:Silberstein} for $\lambda_{0}^t < k \leq \lambda_{0}^0 = b$.
\end{IEEEproof}

\textit{Remark 2.} For the given file size $M=\theta-1$, the authors proved in~\cite{key-12} that FR codes based on regular graphs can attain the bound in \eqref{eq:boundk}. In this paper, we show that FR codes constructed from $t$-designs require the smallest possible reconstruction degree $k$ for those file sizes ranging from $v-t+1$ to $v$.

\section{Tensor Product of FR Codes}

Let $\cC = (P, \cB, \cI)$ be an $(n,\alpha,\theta,\rho)$-FR code and $\cC' = (P', \cB', \cI')$ an $(n',\alpha', \theta', \rho')$-FR code, satisfying the condition that
\begin{equation}
\frac{\alpha}{ \theta} = \frac{\alpha'}{ \theta'}.
\label{eq:grid_condition}
\end{equation}
Denote the blocks in $\cC$ and $\cC'$ by $B_1, B_2, \ldots, B_n$, and $B'_1, B'_2, \ldots, B'_{n'}$, respectively.

We define the {\em tensor product} of $\cC$ and $\cC'$, denoted by $\cC \otimes \cC'$, as the FR code with $\theta \cdot \theta'$ points and $n+n'$ blocks. The points are the pairs in $P\times P'$, and the blocks are given by
\begin{align*}
& B_i\times P', \text{ for } 1\leq i\leq n, \text{ and} \\
& P\times B'_j, \text{ for } 1\leq j\leq n'.
\end{align*}
Notice that the sizes of $B_i\times P'$ and $P\times B'_j$ are $\alpha \theta'$ and $\alpha' \theta$, respectively, and they are equal by the hypothesis in \eqref{eq:grid_condition}. Moreover, we observe that each point in $P\times P'$ appears in exactly $\rho+\rho'$ blocks. Therefore, the tensor product of $\cC$ and $\cC'$ is an FR code with parameters $(n+n', \alpha \theta',\theta \theta', \rho+\rho')$.

\begin{example}
Let $\cC = (P,\cB)$ be the trivial $(g,1,g,1)$-FR code in which each node stores a unique code symbol, i.e., $P=\{1,2,\ldots, g\}$ and $\cB = \{ \{1\},\{2\},\ldots, \{g\}\}$. Then the tensor product $\cC \otimes \cC$ forms a $(2g, g, g^2,2)$-FR code. Specifically, the points are the pairs $(i,j)$ for $i,j\in\{1,2,\ldots, g\}$, and the $2g$ blocks are
\begin{align*}
&\{(i,1),(i,2),\ldots, (i,g)\}, \text{ for } i=1,\ldots, g, \text{ and}\\
&\{(1,j),(2,j),\ldots, (g,j)\}, \text{ for } j=1,\ldots, g.
\end{align*}
This is the same as the $g\times g$ grid code considered in~\cite{key-16}.
\end{example}

\begin{example} Let $\cC = (P,\cB)$ be the trivial $(g,1,g,1)$-FR code as in the previous example. We can take the tensor product $\cC \otimes \cC \otimes \cC$ and obtain a $(3g, g^2, g^3,3)$-FR code. We call this the triple tensor product of $\cC$. The points are the triples $(i,j,\ell)$ for $i,j,\ell \in\{1,2,\ldots, g\}$. The blocks are
\begin{align*}
&\{(i,j,\ell):\, j,\ell \in\{1,\ldots, g\}\}, \text{ for } i=1,\ldots, g, \\
&\{(i,j,\ell):\, i,\ell \in\{1,\ldots, g\}\}, \text{ for } j=1,\ldots, g, \\
&\{(i,j,\ell):\, i,j \in\{1,\ldots, g\}\}, \text{ for } \ell =1,\ldots, g,
\end{align*}
and each block contains $g^2$ points.
\end{example}

We shall list some simple properties about the tensor product of FR codes.

\begin{lem}
For $i=1,2,3$, let $\cC_i$ be an $(n_i, \alpha_i, \theta_i, \rho_i)$-FR code, such that $\alpha_1/\theta_1 = \alpha_2/\theta_2 = \alpha_3/\theta_3$.
\begin{enumerate}
\item $\cC_1 \otimes \cC_2$ and $\cC_2\otimes \cC_1$ are isomorphic FR codes.
\item $(\cC_1 \otimes \cC_2) \otimes \cC_3 = \cC_1 \otimes (\cC_2 \otimes \cC_3)$.
\end{enumerate}
\end{lem}

Moreover, the file size hierarchy of $\cC_1\otimes \cC_2$ can be computed by the following theorem.

\begin{thm}
Let $\cC_i$ be an $(n_i, \alpha_i, \theta_i, \rho_i)$-FR code, for $i=1,2$, such that $\alpha_1/\theta_1 = \alpha_2/\theta_2$. Let $N_k(\cC_1)$ and $N_k(\cC_2)$ be defined as in \eqref{NkC}. We have
\begin{equation}
N_k(\cC_1 \otimes \cC_2) = \max_{\substack{x\in \{0,1,\ldots, n_1\} \\ y\in \{0,1,\ldots, n_2 \} \\ x+y = k }} N_{x}(\cC_1)  N_{y}(\cC_2),
\end{equation}
for $k=1,2,\ldots, n_1+n_2$.
\end{thm}

\begin{IEEEproof}
The incidence matrix of $\cC_1 \otimes \cC_2$ is an $(n_1 + n_2) \times \theta_1 \theta_2$ binary matrix. Without loss of generality, we assume that the first $n_1$ rows correspond to the $n_1$ blocks generated by the $n_1$ blocks of $\cC_1$ and the other $n_2$ rows correspond to the $n_2$ blocks obtained by the $n_2$ blocks of $\cC_2$. Consider now we have $k=x+y$ blocks of $\cC_1 \otimes \cC_2$, among which $x$ blocks are taken from~the first $n_1$ rows and $y$ blocks are from the last $n_2$ rows.

We first consider the $x \times \theta_1 \theta_2$ submatrix corresponding to the $x$ blocks. Based on the tensor product method, we have that the maximum integer $\xi$ such that there exists an $x\times \xi$ all-zero submatrix in the $x \times \theta_1 \theta_2$ matrix is $(\theta_1-M_{x}(\cC_1))\theta_2$, i.e., $\xi = N_{x}(\cC_1)\theta_2$. By jointly considering the $y$ rows from the last $n_2$ rows, we obtain that the maximum integer $\zeta$ such that there exists a $k\times \zeta$ all-zero submatrix in the $k \times \theta_1 \theta_2$ matrix is $N_{x}(\cC_1)(\theta_2-M_{x}(\cC_2)) = N_{x}(\cC_1) N_{y}(\cC_2)$, which completes the proof.
\end{IEEEproof}

\begin{figure*}
$${
\begin{tabular}{|c|c|c|c|c|} \hline
$1, 2, 3, 16$ & $1, 2, 3, 17$ & $1, 2, 3, 18$ & $1, 2, 3, 19$ & $1, 2, 3, 20$ \\ \hline
$4, 5, 6, 16$ & $4, 5, 6, 17$ & $4, 5, 6, 18$ & $4, 5, 6, 19$ & $4, 5, 6, 20$ \\ \hline
$7, 8, 9, 16$ & $7, 8, 9, 17$ & $7, 8, 9, 18$ & $7, 8, 9, 19$ & $7, 8, 9, 20$ \\ \hline
$10, 11, 12, 16$ & $10, 11, 12, 17$ & $10, 11, 12, 18$ & $10, 11, 12, 19$ & $10, 11, 12, 20$ \\ \hline
$13, 14, 15, 16$ & $13, 14, 15, 17$ & $13, 14, 15, 18$ & $13, 14, 15, 19$ & $13, 14, 15, 20$ \\ \hline
\end{tabular}
}
$$
\caption{A $(5,3,1)$-GFR code. Each entry corresponds to a distinct storage node, and the numbers in an entry correspond to the coded packets stored in the storage node.}
\label{fig:531}
\end{figure*}

\begin{cor}
Let $s$ and $e_1,\ldots, e_s$ be  positive integers. For $i=1,2,\ldots, s$, let $\cC_i$ be an $(n_i, \alpha_i, \theta_i, \rho_i)$-FR code,  such that $\alpha_i/\theta_i$ is equal to a constant $c$ for all $i$. Let $\cC_i^{e_i}$ be the FR code obtained from $\cC_i$ by repeating each of the blocks in $\cC_i$ $e_i$-fold. Then $\cC_1^{e_1} \otimes \cC_2^{e_2} \otimes \cdots \otimes \cC_s^{e_s}$ is an FR code with parameters
$$
(n,\alpha, \theta, \rho) = \big( \sum_{i=1}^s e_i n_i , c \prod_{i=1}^s \theta_i , \prod_{i=1}^s \theta_i, \sum_{i=1}^s \rho_i e_i\big)
$$
and the file size hierarchy can be determined by
$$
N_k(\cC_1^{e_1} \otimes \cC_2^{e_2} \otimes \cdots \otimes \cC_s^{e_s})
$$
\begin{equation}
= \max_{ \substack{x_i\in \{0,1,\ldots, n_i\}, 1\leq i\leq s \\ e_1x_1+\cdots + e_s x_s = k }} N_{x_1}(\cC_1) N_{x_2}(\cC_2)  \cdots N_{x_s}(\cC_s),
\end{equation}
for $k=1,2,\ldots, \sum_{i=1}^s e_i n_i$.
\label{thm:grid}
\end{cor}

\begin{example}
Let $g$ and $s$ be integers  larger than or equal to 2. Let $\mathcal{G}$ denote the trivial $(g,1,g,1)$-FR code with the $g\times g$ identity matrix as the incidence matrix. For positive integers $\alpha_1,\ldots, \alpha_s$, consider the FR code
$$
\cC = (\mathcal{G}^{\alpha_1} \otimes \mathcal{G}^{\alpha_2} \otimes \cdots  \otimes  \mathcal{G}^{\alpha_s})^t,
$$
and denote it by a $(g,\alpha_1,\ldots, \alpha_s)$-GFR code. The resulting FR code has parameters
$$
(n,\alpha,\theta,\rho) = (g^s, \sum_{i=1}^{s}\alpha_{i}, g\sum_{i=1}^{s}\alpha_{i}, g^{s-1}).
$$
Fig.~\ref{fig:531} shows how to distribute $20$ coded packets across $25$ storage nodes by a $(5,3,1)$-GFR~code. Since the file size hierarchy of $\mathcal{G}$ is simply given by
$$
M_k(\mathcal{G}) = k, \text{ for } k=1,2,\ldots, g,
$$
we can apply Theorem~\ref{thm:duality} and Corollary~\ref{thm:grid} and obtain the file size hierarchy of the $(5,3,1)$-GFR code $\cC$ as
$$
M_k(\cC) = \begin{cases}
20, & \text{for } k = 21, 22, 23, 24, 25,\\
17, & \text{for } k = 17,18,19,20,\\
16, & \text{for } k = 16,\\
14, & \text{for } k = 13, 14, 15,\\
13, & \text{for } k = 11, 12,\\
11, & \text{for } k = 9, 10,\\
10, & \text{for } k = 7, 8,\\
9, & \text{for } k = 6,\\
8, & \text{for } k = 5,\\
7, & \text{for } k = 4,\\
6, & \text{for } k = 3,\\
5, & \text{for } k = 2,\\
4, & \text{for } k = 1.
\end{cases}
$$
\end{example}

\textit{Remark 3.} Olmez and Ramamoorthy~\cite{key-16} presented the Kronecker product technique for combining two FR codes, where they analyzed the supported file size and failure resilience of the resulting code for some special scenarios. In this paper, we study the tensor product~of two FR codes, and characterize the file size hierarchy of the resulting product code.

\section{Conclusion}

Determining the supported file size $M_k(\cC)$ of an FR code $\cC$ is a challenge task in the domain of FR codes. In this paper, we provide an alternative viewpoint by considering the ``complementary supported file size'', which is defined as the total number of distinct packets in $\cC$ minus $M_k(\cC)$. Specifically, we first establish a close relationship between the file size hierarchy of an FR code and its dual code. Based on the relationship, we derive a dual bound on the supported file size, which is tighter than the existing upper bounds in some cases. From the dual perspective, we prove that the supported file size of $t$-design based FR codes is optimal when the size of the stored file is sufficiently large. We also propose the tensor product method for combining two FR codes. The hierarchy of complementary supported file size of the resulting product code can be expressed as a kind of ``convolution'' of those of the component codes. Although we focus on FR codes in which each storage node contains the same number of packets and each packet is stored in the same number of nodes, the basic idea can also be generalized beyond this symmetric case. Extension to heterogeneous FR codes is interesting for future exploration.

\end{document}